\documentclass[11pt]{article}
\usepackage{amsmath}
\usepackage{amssymb}
\usepackage{epsfig}
\usepackage[bf,small]{caption}
\def\tr{{\rm tr}\ }

\begin{document}


\begin{titlepage}

\renewcommand{\thefootnote}{\alph{footnote}}
\vspace*{-3.cm}
\begin{flushright}

\end{flushright}

\vspace*{0.3in}

\renewcommand{\thefootnote}{\fnsymbol{footnote}}
\setcounter{footnote}{-1}

{\begin{center} {\Large\bf The  Symplectic-Orthogonal Penner Models }

\end{center}}
\renewcommand{\thefootnote}{\alph{footnote}}

\vspace*{.8cm} {\begin{center} {\large{\sc
                Mohammad~Dalabeeh$^a$ and Noureddine Chair$^{b ,1}$\footnote{$^{1}$The corresponding author.}
                }}
\end{center}}
\vspace*{0cm} {\it
\begin{center}
 $^{a,b}$Physics Department,
 University of Jordan, Amman, Jordan

$^{a}$Email:maaldlabyh05@sci.ju.edu.jo
$^{b}$Email:n.chair@ju.edu.jo
\end{center} }

\vspace*{1.5cm}

{\Large \bf
\begin{center} Abstract\end{center} }
The generating function for the orbifold Euler characteristic of the moduli space of real algebraic curves of genus $2g$ (locally orientable surfaces) with $n$ marked points $\chi^r(\mathfrak{M}_{2g,n})$, is identified with a simple formula. It is shown that the free energy in the continuum limit of both the symplectic and the orthogonal Penner models are almost identical, with the structure $F^{SP/SO}(\mu)=\frac{1}{2}F(\mu)\mp F^{NO}(\mu)$, where $F(\mu)$ is the Penner free energy and $F^{NO}(\mu)$  is the free energy contributions from the non-orientable surfaces. Both of these models have the same critical  point as the Penner model. 
\end{titlepage}

\renewcommand{\thefootnote}{\arabic{footnote}}
\setcounter{footnote}{0}
\newpage

\section{INTRODUCTION}
\  The formula for the orbifold Euler characteristic of the moduli space of real algebraic curves with $n$ marked points (punctures) was first obtained by  Chekhov and Zabrodin,  using the orthogonal matrix model technique (skew orthogonal polynomials)\cite{l.chekhov}.   Goulden, Harer and Jackson \cite{jakson} 
obtained  the same  formula, as a special case of a more general formula determined from the parametrized polynomial $\xi_{g}^n(\gamma)$, when $\gamma =1/2$, where $g$ is the genus of the algebraic curves and $n$ is the number of punctures. These authors used  real symmetric matrix integrals  rather than hermitian matrix integrals which are known to give the orbifold Euler characteristic of the moduli space of complex algebraic curves (orientable surfaces) \cite{penner,harer}. It was shown in \cite{jakson} that the real symmetric matrix integral gives rise to two different  orbifold Euler characteristics, namely: When $g$ is odd, it is  the ordinary orbifold Euler characteristic of the moduli space of complex algebraic curves with $n$ marked points. However, when $g$ is even they obtained  the orbifold Euler characteristic of the moduli space of real algebraic curves with $n$ marked points. This contributions is due to the non-orientable surfaces.
This model is called the orthogonal Penner model. Mulase and Waldron \cite{mulase1} generalized the Penner model to the symplectic Penner model \cite{mulase1} and found that the orthogonal Penner model almost coincides with the symplectic Penner model, the difference being in that the  matrix  size of the former is doubled and an overall sign difference  appears in  the non-orientable surfaces contributions. This was also observed for a general potential potential $V(x)$ by Chekhov and Eynard \cite{eynard}\footnote{The  formula relating the symplectic and the orthogonal matrix models for a general potential $V(x)$ reads:
\begin{equation*}
-\log\int\left|\Delta(\lambda)\right|^{2\alpha} e^{-N\sqrt\alpha\sum_{i=1}^{N}V(\lambda_{i})}=\sum_{g=0}^{\infty}\sum_{k=0}^{\infty}N^{2-2g-k}\left(\sqrt{\alpha}-\sqrt{\alpha^{-1}}\right)^kF_{g,k},
\end{equation*}
where $F_{g,k}$ are the corresponding correction to the free energy. Here $\alpha=1/2$ for the orthogonal matrices and $\alpha=2$ for the symplectic ones.}.   Our goals in this paper are two fold: First, We identify the generating function for the orbifold Euler characteristic of the moduli space of real  algebraic curves when $g$ is even which turns out to be a simple formula,  this is done in section $2$. The second goal is to obtain the continuum limit of the symplectic-orthogonal Penner models. This will be carried out in section $3$. Like the ordinary Penner model \cite{dvafa,chair1}, we show that  the free energy of the symplectic-orthogonal Penner models in the continuum limit are related to the orbifold Euler characteristic of the moduli space without punctures. Both have the same critical point $t=1$. This is the same critical point of  the ordinary Penner model \cite{chair1}.  

\section{The Generating Function of the Penner  Symplectic-Orthogonal Matrix Models }

\ The Penner model for hermitian matrix integral provides an effective tool to compute the orbifold Euler characteristic of the moduli space of smooth algebraic curves defined over $\mathbb{C}$  with an arbitrary number of marked points \cite{harer}.
 While the asymptotic expansion of the Penner model for real symmetric and quaternionic    matrix integral is the generating function of the orbifold Euler characteristic of the moduli space of real algebraic curves \cite{jakson}.
 
 \  The asymptotic expansion of the free energy of the Penner model is  constructed from the logarithm of Gaussian integral of self-adjoint matrices. {The Gaussian  integral of self-adjoint matrices  as a function of the coupling constants 
$t = (t_1, t_2, t_3,  \ldots)$, and the size of the matrix $N$, is given by
\begin{equation}
Z^{(2\alpha)}(t,N)=\frac{\int[dX]_{_{(2\alpha)}}\exp\bigl(-\frac{1}{4} \tr X^2+\sum_{j=1}^{\infty}\frac{t_j}{2j}\tr X^{j}\bigr)}{\int[dX]_{_{(2\alpha)}}\exp\bigl(-\frac{1}{4} \tr X^2\bigr)},
\end{equation}where the matrix variable $X$ is constructed from $N\times N$ real symmetric and complex anti symmetric matrices.} Using the Penner substitution given by
 \begin{equation}
t_j=-(\sqrt{t})^{j-2},\quad j=3,4,\dots ,2m \quad \text{with}\quad t_1=t_2=0,
\end{equation} 
where $\sqrt{t}$ is defined for $Re(t)>0$, then the free energy of the Penner model reads   
\begin{equation}
F(t,N,\alpha)=\lim_{m\rightarrow\infty}\log\left(\frac{\int[dX]_{_{(2\alpha)}}\exp\bigl(-\sum_{j=2}^{2m}\frac{t^{j/2-1}}{j}\tr X^{j}\bigr)}{\int[dX]_{_{(2\alpha)}}\exp\bigl(-\frac{1}{2} \tr X^2\bigr)}\right),
\end{equation}
where $\alpha$ takes the values $\frac{1}{2},1$ or $2$ depending on whether we study orthogonal ,unitary or symplectic ensembles   respectively. Also note that the sum over $j$ in the above equation 
contains  the quadratic term i.e $j=2$,  as well as the interaction terms starting from $j\geq3$ with  conventional normalization\footnote{ To obtain a conventional normalization, we rescale $X\rightarrow 2^{1/2} X$ in equation (1) and absorb all but one power of $2$ in the couplings $t$.} .  This integral is  computable since the matrix variable $X$ is diagonalizable, ${X\rightarrow diag(\lambda_1,\lambda_2,\dots,\lambda_N)}$ where $\lambda$'s  are the eigenvalues of the matrix $X$, therefore the integral becomes
\begin{equation}
\label{mas}
F(t,N,\alpha)=\lim_{m\rightarrow\infty}
\log\left(\frac{
\int_{{\mathbb R}^N}\Delta^{2\alpha}(\lambda)\prod_{i=1}^N
\exp\Big(-\sum_{j=2}^{2m} \frac{t^{j/2-1}}{j}\ \lambda_i^j\Big) d\lambda_i}
{\int_{{\mathbb R}^N}\Delta^{2\alpha}(\lambda)\prod_{i=1}^N
\exp\Big(- \frac{\lambda_i ^2}{2}\Big) d\lambda_i}
\right),
\end{equation}
where  $\Delta(\lambda)=\prod_{i<j}(\lambda_i-\lambda_j)$, is the Vandermonde determinant.

Using the Selberg integration formula, Stirling's formula for $\Gamma(1/t)$ and the asymptotic analysis of \cite{Mulase95},
the above integral can be evaluated for integer values of $\alpha$, and the result is 
\begin{equation}
\begin{split}
\label{alpha}
F(t,N,\alpha)=& \sum^{\infty}_{m=1}{\frac{B_{2m}}{2m(2m-1)}Nt^{2m-1}}\\&+\sum^{\infty}_{m=1}\sum^{N-1}_{i=0}\sum^{\alpha}_{j=1}{(-1)^{m-1}}\frac{1}{m} (N-1-i)(i\alpha+j)^{m}t^{m},
\end{split}
\end{equation}
where $B_{2n}$ are the Bernoulli numbers. 
The free energy of the original Penner model (integral over hermitian matrices) is obtained   by setting $\alpha=1$ in (\ref{alpha})  
\begin{equation}
\label{penner or}
F(t,N,1)
=\sum_{\substack{g\ge 0, n>0\\2-2g-n<0}}
\frac{(2g+n-3)!(2g-1)}{(2g)!n!}B_{2g} N^n (-t)^{2g+n-2},
\end{equation}
identifying $n$ with the number of faces $f_\Gamma$, and $g$ as the genus of the triangulated Riemann  surfaces. The free energy $F(t,N,1)$ is considered  the generating function of the orbifold  Euler characteristic $\chi^c(\mathfrak M_{g,n})$ of the moduli space of Riemann surfaces   of genus $g$ and $n$ punctures given by   
\begin{equation}
\chi^c(\mathfrak M_{g,n})=(-1)^n\frac{(2g+n-3)!(2g-1)}{(2g)!n!}B_{2g}. 
\end{equation}
The partition function of the Penner model  can be evaluated using the orthogonal polynomial technique \cite{penner}, given by 
 \begin{equation}
 \label{Penner}
 \ \Bigr(\frac{\sqrt{2\pi t}(et)^{-(t^{-1})}}{\Gamma(\frac{1}{t})}\Bigl)^{N}\prod_{p=1}^{N}(1+pt)^{(N-p)}.
 \end{equation}
By setting  $\alpha =2$ in equation (\ref{alpha}) one may show that the free energy of the symplectic Penner model \cite{mulase1} reads:
\begin{equation}
\begin{split}
\label{penner g}
F(t,N,2)=&\frac{1}{2}\sum_{\substack{g\ge 0, n>0\\2-2g-n<0}}
\frac{(2g+n-3)!(2g-1)}{(2g)!n!}B_{2g} (2N)^n (-t)^{2g+n-2}\\ &-\frac{1}{2}\sum_{\substack{q\ge 0,n>0\\ 1-2q-n<0}}
 \frac{(2q+n-2)!(2^{2q-1}-1)}{(2q)!\;n!}B_{2q}
(2N)^{n} (-t)^{2q+n-1}.
\end{split}
\end{equation}
As we can see from the above equation, the  first term is half  the free energy of the   Penner model with the size of the matrix doubled i.e. $\frac{1}{2}F(t,2N,1)$. This term comes from the contributions of the  orientable surfaces,   while  the  second term is the  contributions from  the  non-orientable surfaces of even genus $g=2q$ with $n$ marked points, where the term
\begin{equation}  
\chi^r(\mathfrak M_{2q,n}):=(-1)^n\frac{1}{2}\frac{(2q+n-2)!(2^{2q-1}-1)}{(2q)!\;n!}B_{2q},
\end{equation}
is the orbifold Euler characteristic of the moduli space of smooth real algebraic curves of genus $g=2q$ with $n$ marked points \cite{jakson}. As a consequence, the free energy of the symplectic Penner model is the generating function of two different  orbifold Euler characteristics namely $\chi^c(\mathfrak M_{2q,n})$ and $\chi^r(\mathfrak M_{g,n})$.  The first one  is generated by    
\begin{equation}
\label{symPenner}
\log\Bigr[(\frac{\sqrt{2\pi t}(et)^{-(t^{-1})}}{\Gamma(\frac{1}{t})})^{2N}\prod_{p=1}^{2N}(1+pt)^{(2N-p)}\Bigl].
\end{equation}
One of our objectives in this paper is to  obtain  the generating function for  $\chi^r(\mathfrak M_{2q,n})$. In so doing,  we begin by rewriting the asymptotic expansion of the symplectic Penner model given in equation (\ref{alpha})  as  two separate  generating functions in which  the first  is given by  equation (\ref{symPenner}) and the second one  is identified with the generating function of the orbifold Euler characteristic of smooth real algebraic curves of genus $g=2q$ with $n$ marked points $\chi^r(\mathfrak M_{2q,n})$. From equation (\ref{alpha}), the symplectic Penner model is 
\begin{equation}
F(t,N,2)=\sum^{\infty}_{m=1}{\frac{B_{2m}}{2m(2m-1)}Nt^{2m-1}}+\sum^{\infty}_{m=1}\sum^{N-1}_{i=0}\sum^{2}_{j=1}{(-1)^{m-1}}\frac{1}{m} (N-1-i)(2i+j)^{m}t^{m}.
\end{equation}
Now, expanding the sum over $j$ and using the Maclaurin  series expansion for the sum over $m$, 
 the asymptotic expansion of the symplectic Penner model  becomes
\begin{equation}
\begin{split}
F(t,N,2)=&\sum^{\infty}_{m=1}{\frac{B_{2m}}{2m(2m-1)}Nt^{2m-1}}\\&+\sum^{N-1}_{i=0}(N-1-i)\bigr[\log(1+(2i+1)t)
+\log(1+(2i+2)t)\bigl],
\end{split}
\end{equation}
the summation over $i$  can be written as follows  
\begin{equation}
\begin{split}
F(t,N,2)=&\sum^{\infty}_{m=1}{\frac{B_{2m}}{2m(2m-1)}Nt^{2m-1}}\\
&+\sum_{p_{odd}=1}^{2N-1}\frac{1}{2}(2N-1-p)\log(1+pt)+\sum_{p_{even}=2}^{2N}\frac{1}{2}(2N-p)\log(1+pt),
\end{split}
\end{equation}
Combining the similar expressions  for both odd and even summations, the restrictions may be  lifted and gives $\small{\sum_{p=1}^{2N}\frac{1}{2}(2N-p)\log(1+pt)}$, with one remaining restricted term namely, $\small{\sum_{p=odd}^{2N-1}\log(1+pt)^{-\frac{1}{2}}}$.  The free energy expression becomes  
\begin{equation}
F(t,N,2)=\sum^{\infty}_{m=1}{\frac{B_{2m}}{2m(2m-1)}Nt^{2m-1}}+\log\prod_{p=1}^{2N}(1+pt)^{\frac{1}{2}{(2N-p)}}-\log\prod_{p_{odd}=1}^{2N-1}(1+pt)^{\frac{1}{2}}.
\end{equation}   
We claim that the last term  in which $p$ is restricted to be odd  is responsible for the  contributions coming  from graphs drawn on non-orientable Riemann surfaces. In order to see this, we point out  that the combination of the first and the second  terms  is half of the  Penner model,  with the size of the matrix doubled. Using the Stirling formula for  the logarithm of $\Gamma(\frac{1}{t})$ 
\begin{equation}
\log(\Gamma(\frac{1}{t}))=-\frac{1}{t}\log t -\frac{1}{t}+\frac{1}{2}\log t+\sum^{\infty}_{m=1}{\frac{B_{2m}}{2m(2m-1)}t^{2m-1}}+\text{const}
\end{equation}
 the symplectic Penner model finally reads:
\begin{equation}
\label{penner symplectic}
F(t,N,2)=\frac{1}{2}\log\Bigr[(\frac{\sqrt{2\pi t}(et)^{-(t^{-1})}}{\Gamma(\frac{1}{t})})^{2N}\prod_{p=1}^{2N}(1+pt)^{(2N-p)}\Bigl]-\frac{1}{2}\log\prod_{p_{odd}=1}^{2N-1}(1+pt).
\end{equation}
Writing the first  term in equation (\ref{penner symplectic}) as $\frac{1}{2}F(t,2N,1)$, Then the symplectic Penner model  reads: 
\begin{equation}
F(t,N,2)=\frac{1}{2}F(t,2N,1)-\frac{1}{2}\log\prod_{p_{odd}=1}^{2N-1}(1+pt).
\end{equation}
In order to see that the second term is the generating function of the orbifold Euler characteristic $\chi^r(\mathfrak M_{g,n})$,  we first expand the latter,
\begin{equation}
\label{prod odd}
\log\prod_{p_{odd}=1}^{2N-1}(1+pt)
 = \sum_{m=1}^{\infty}{(-1)^{m-1}}\frac{1}{m}t^m\sum_{p_{odd}=1}^{2N-1}p^{m}.
 \end{equation}
 Now, the restricted sum in the above equation may be lifted using 
\begin{equation}
\sum_{p_{odd}=1}^{2N-1}p^{m}=\sum_{p=1}^{2N}p^{m}-\sum_{p=1}^{N}(2p)^{m},
\end{equation}
and from the power sum formula  
\begin{eqnarray}
\sum_{p=1}^{N}{p^{m}}=\frac{N^{m+1}}{m+1}+\frac{N_{m}}{2}+\sum_{k=1}^{[\frac{m}{2}]}\binom{m}{2k-1}\frac{B_{2k}}{2k}N^{m+1-2k},
\end{eqnarray}
equation (\ref{prod odd}) can be written as, 
\begin{equation}
\begin{split}
\log\prod_{p_{odd}=1}^{2N-1}(1+pt)=&
\sum_{m=1}^{\infty}{(-1)^{m-1}}\frac{1}{m}\frac{N^{m+1}}{m+1}(2t)^m\\- &\sum_{m=1}^{\infty}\sum_{k=1}^{[\frac{m}{2}]}\frac{1}{m}\binom{m}{2k-1}\frac{B_{2k}}{2k}N^{m+1-2k}(2^{2k-1}-1)(2t)^{m}.
\end{split}
\end{equation}
Combining the above two sums we get 
\begin{eqnarray}
\log\prod_{p_{odd}=1}^{2N-1}(1+pt)=\sum_{m=1}^{\infty}\sum_{k=0}^{[\frac{m}{2}]}\frac{(m-1)!(2^{2k-1}-1)}{(2k)!(m+1-2k)!}B_{2k}(2N)^{m+1-2k}(-t)^{m},
\end{eqnarray}
replacing  $ m \text{\space by\space} 2k+n-1$ and $k \text{\space by\space}q$ in the above equation then, 
\begin{equation}
\begin{split}
\label{penner nor}
\log\prod_{p_{odd}=1}^{2N-1}(1+pt)^{-\frac{1}{2}}=&-\frac{1}{2}
\sum_{\substack{q\ge 0,n>0\\ 1-2q-n<0}}
 \frac{(2q+n-2)!(2^{2q-1}-1)}{(2q)!\;n!}B_{2q}
(2N)^{n} (-t)^{2q+n-1}.
\end{split} 
\end{equation}
Therefore, this shows  that   $\log\prod_{p_{odd}=1}^{2N-1}(1+pt)^\frac{1}{2}$  is indeed the generating function of the  orbifold Euler characteristic of the moduli space of smooth real algebraic curves of genus $2q$ with $n$ marked points.
From Mulase and Waldron \cite{mulase1}, the generating function of the orthogonal Penner model $F(2t,2N,2)$, is given by the following expression  
 \begin{equation}
 \label{orthogonal}
\begin{split}
F(2t,2N,2)=&\frac{1}{2}\sum_{\substack{g\ge 0, n>0\\2-2g-n<0}}
\frac{(2g+n-3)!(2g-1)}{(2g)!n!}B_{2g} (2N)^n (-t)^{2g+n-2}\\ &+\frac{1}{2}\sum_{\substack{q\ge 0,n>0\\ 1-2q-n<0}}
 \frac{(2q+n-2)!(2^{2q-1}-1)}{(2q)!\;n!}B_{2q}
(2N)^{n} (-t)^{2q+n-1}.
\end{split}
\end{equation}
 Therefore, we  conclude  that the generating function for  the non-orientable surfaces contributions to the  orthogonal Penner model is
\begin{equation} 
\log\prod_{p_{odd}=1}^{2N-1}(1+pt)^{\frac{1}{2}}.
\end{equation}
Note that, the structure of the symplectic and the orthogonal Penner matrix models equations (\ref{penner g}),(\ref{orthogonal}), is reminiscent of the $SO(N)$ and $Sp(N)$ Chern-Simons gauge theory on $S^3$ \cite{vafa}.

\section{The Continuum Limit of  Symplectic-Orthogonal Penner  Matrix Models}

It is clear from our previous section on the generating functions of the symplectic Penner model, that the continuum limit is obtained by adding
 half the continuum limit of the ordinary Penner model $F(t,N,1)$, to the continuum limit of the non orientable surfaces contributions.   
 The  continuum limit of  the ordinary Penner model was obtained in \cite{dvafa} and \cite{chair1}, where   they found  that the Penner free 
energy  is the generating function of the orbifold Euler characteristic without punctures $\chi^c(\mathfrak M_{g,0})$. This means that  graphs with $n$ punctures  are not 
present  in the continuum limit of the free energy. The continuum limit of the free energy of the ordinary Penner model  $F(t,N,1)$  is given by
\begin{equation}
F(\mu)=\frac{1}{2}\mu^2 \log\mu
-\frac{1}{12}\log\mu+\sum_{g\geq2}\frac{1}{(2g-2)}\frac{B_{2g}}{2g}\mu^{2-2g},
\end{equation}
where $\frac{1}{2}\mu^2 \log\mu$ and $\frac{1}{12}\log\mu$
are the sphere and the torus contributions to the free energy respectively. While $\frac{B_{2g}}{2g(2g-2)}$ is the orbifold Euler
 characteristic without punctures $\chi^c{(\mathfrak M_{g, 0})}$. Therefore, in order to obtain the  continuum limit for the symplectic Penner model one need only  focus  on the non-orientable contributions part. First, we rewrite equation (\ref{prod odd}) as follows;  
\begin{equation}
\log\prod_{p_{odd}=1}^{2N-1}(1+pt)^{-\frac{1}{2}}=-\frac{1}{2}\bigr[\sum_{p=1}^{2N}\log(1+pt)-\sum_{p=1}^{N}\log(1+2pt)\bigl],
\end{equation}
  using the Euler Maclaurin formula 
\begin{equation}
\begin{split}
  \sum_{p=1}^{2N}\log(1+pt) =&\frac{1}{2}\bigr[f(2N)-f(1)\bigr] + \int_{1}^{2N}\log(1+xt)dx \\ &+\sum_{k=1}^{\infty}\frac{B_{2k}}{(2k)!}\bigr[f^{(2k-1)}(1)-f^{(2k-1)}(2N)\bigl],
  \end{split}
  \end{equation} 
   where $f$ stands for $\log(1+xt)$, and $f^{(k)}$  is the $kth$ derivative of  $f$,  we obtain 
\begin{equation}
\begin{split}
\label{sympcon}
\log\prod_{p_{odd}=1}^{2N-1}(1+pt)^{-\frac{1}{2}}&=\frac{1}{4}\log\frac{1+t}{1+2t}+\frac{1}{2}\sum_{k=1}^{\infty}\frac{B_{2k}}{(2k)(2k-1)}(2^{2k-1}-1)\Bigr(\frac{t}{1+2Nt}\Bigl)^{2k-1}\\&+\frac{1}{2}\sum_{k=1}^{\infty}\frac{B_{2k}}{(2k)(2k-1)}\Bigr[(\frac{t}{1+t})^{2k-1}-(\frac{2t}{1+2t})^{2k-1}\Bigl]-\frac{1}{2}\bigr[(1+2t)\log(1+2t)\\&\frac{1}{2t}(1+2Nt)\log(1+2Nt)-\frac{1}{t}(1+t)\log(1+t)-N\bigl].
\end{split}
\end{equation}
The continuum limit of the non-orientable contributions in equation (\ref{sympcon}) is obtained by first making the natural scaling $t\longrightarrow -t/2N$, then set $\mu=2N(1-t)$, and let  $N\rightarrow\infty$, $t\rightarrow1$\footnote{This is the same as in the ordinary Penner model \cite{chair1}} such that $\mu$ is kept fixed (double scaling). Doing so, we get: 
\begin{equation}
\label{cont}
F^{NO}(\mu)=\frac{1}{4}\mu\log\mu-\sum_{k\geq1}^{\infty}\frac{(2^{2k-1}-1)}{(2k-1)}\frac{B_{2k}}{4k}\mu^{1-2k},
\end{equation}
where $ F^{NO}(\mu)$  is the non-orientable surfaces contributions to the continuum limit of the symplectic   Penner free energy.
 The term  $\frac{1}{4}\mu\log\mu $ is the $g=0$ contribution and $\frac{(2^{2k-1}-1)}{(2k-1)}\frac{B_{2k}}{4k}$  is the orbifold Euler characteristic of
 the moduli space of smooth real algebraic curves without marked points  $\chi^r{(\mathfrak M_{g, 0})}$.
 
Alternatively, the continuum limit may be obtained by summing over all punctures using   equation(\ref{penner nor}). The $g=0$ contribution corresponds to the following sum
\begin{equation}
\frac{1}{4}
\sum_{\substack{ n>1}}
 \frac{1}{\;n(n-1)}
(2N)^{n} (-t)^{n-1}.
\end{equation}
Making the substitution $t\rightarrow \frac{-t}{2N}$, then the  above  sum reads
\begin{equation}
\frac{N}{2}
\sum_{\substack{ n>1}}
 \frac{1}{\;n(n-1)} (t)^{n-1}=\frac{N}{2}\Bigr{[}\sum_{n=2}^{\infty}\frac{(t)^{n-1}}{n-1}-\sum_{n=2}^{\infty}\frac{(t)^{n-1}}{n}\Bigl{]}
\end{equation}
which is equivalent to 
\begin{equation}
\frac{N}{2}\Bigr{[}1+(\frac{1-t}{t})\log(1-t)\Bigl{]}.
\end{equation}
 Then, taking the continuum limit, one can  reproduce the  $g=0$ contribution.
Similarly,  for the  higher genus $q\geq 1 $, upon using the identity
\begin{equation}
\nonumber
\frac{d^n}{dt^n}(1-t)^{-2q}=\frac{(2q+n-1)!}{(2q-1)!}(1-t)^{-2q-n},
\end{equation} 
 the sum over punctures gives   
\begin{equation}
-\sum_{q\geq 1}\frac{(2^{2q-1}-1)}{(2q-1)}\frac{B_{2q}}{4q}\Bigl{(}\frac{2N(1-t)}{t}\Bigr{)}^{1-2q},
\end{equation} 
which in turn implies   the continuum limit given by  equation (\ref{cont}).

\   Finally, the free energy of the symplectic Penner model in the continuum limit  reads;
\begin{equation}
\label{finalsym}
F(\mu)=\frac{1}{4}\mu^2 \log\mu-\frac{1}{4}\mu\log\mu
+\frac{1}{24}\log\mu+\frac{1}{2}\sum_{g\geq2}\frac{1}{(2g-2)}\frac{B_{2g}}{2g}\mu^{2-2g}-\sum_{k\geq1}^{\infty}\frac{(2^{2k-1}-1)}{(2k-1)}\frac{B_{2k}}{4k}\mu^{1-2k}.
\end{equation}
Therefore,  this  free energy  is related to the orbifold Euler characteristic without punctures. However, differentiating $F(\mu)$ with respect to  $\mu$, $n$-times provided that $n\geq 3$, brings back the  punctures on  the Riemann surfaces, i.e. we obtain equation (\ref{penner g}). 
Similarly,    the continuum limit of  the orthogonal Penner model can be obtained using  equation (\ref{orthogonal}). Alternatively,  this would be equivalent to change the sign for the non-orientable contributions given by equation (\ref{finalsym}). 

\  It is interesting to note the  connection  between the non-orientable contributions and   $D=1$ string theory \cite{d=1}. This connection can be easily seen  if we make the Wick rotation  $\mu \rightarrow i \mu$ in the density of states $\pi\rho(\mu_F)$ given by the following expression   
\begin{equation}
\nonumber
\pi\rho(\mu_{F})=\frac{1}{2}\Biggr[-\ln\mu +\sum_{m=1}^{\infty}(2^{2m-1}-1)\frac{B_{2m}}{m}\frac{1}{\mu^{2m}}\Biggl],
\end{equation}
integrating the Wick rotated density of states with respect to $\mu$, then   we obtain the non-orientable contributions for the free energy of the symplectic Penner model given by (\ref{cont}).

\    In summary,   the generating function formula for the orbifold  Euler characteristic,  when $g$ is even, is identified and given by the simple formula $\log\prod_{p_{odd}=1}^{2N-1}(1+pt)^{\frac{1}{2}}$. Moreover,  the continuum limits of the symplectic-orthogonal Penner models are obtained and both share    the same critical point $t=1$ with   the Penner model \cite{chair1}. 
The second author found recently that the generating function for $\chi^c{(\mathfrak M_{g, n})}$ is always present \cite{submitted} in all generating functions for the virtual Euler characteristics of moduli spaces \cite{jakson}, as  yet unidentified, meaning that  the orientability is present in all  these moduli spaces. 
\bibliographystyle{phaip}

\end{document}